\documentclass[sigconf]{acmart}

\usepackage{makecell}
\usepackage{listings}
\usepackage{color}
\usepackage{graphicx}
\usepackage{xspace}
\usepackage{multicol}
\usepackage{booktabs}
\usepackage{multirow}
\usepackage{amsmath}
\usepackage{algorithm}
\usepackage{algorithmicx}
\usepackage{algpseudocode}
\usepackage{subcaption}
\usepackage{balance}
\usepackage{todonotes}
\usepackage{nopageno}
\usepackage[T1]{fontenc}
\usepackage{fancyhdr}

\presetkeys{todonotes}{inline}{}

\graphicspath{{figures/}}

\definecolor{dkgreen}{rgb}{0,0.6,0}
\definecolor{gray}{rgb}{0.5,0.5,0.5}
\definecolor{mauve}{rgb}{0.58,0,0.82}

\newcommand{\gadgetone}{\textsc{C-gadget}\xspace}

\newcommand{\gadgettwo}{\textsc{L-gadget}\xspace}

\newcommand{\gadgetthree}{\textsc{E-gadget}\xspace}

\newcommand{\eg}{\textit{e.g., }}
\newcommand{\ie}{\textit{i.e., }}
\newcommand{\etal}{\textit{et al. }}
\newcommand{\etc}{\textit{etc.}}
\newcommand{\na}{TROP}
\newcommand{\name}{Typed ROP (TROP) }

\begin{document}

\copyrightyear{2018} 
\acmYear{2018} 
\setcopyright{acmcopyright}
\acmConference[ACSAC '18]{2018 Annual Computer Security Applications Conference}{December 3--7, 2018}{San Juan, PR, USA}
\acmBooktitle{2018 Annual Computer Security Applications Conference (ACSAC '18), December 3--7, 2018, San Juan, PR, USA}
\acmPrice{15.00}
\acmDOI{10.1145/3274694.3274739}
\acmISBN{978-1-4503-6569-7/18/12}

\title{On the Effectiveness of Type-based Control Flow Integrity}

\author{Reza Mirzazade farkhani}
\affiliation{\institution{Northeastern University}}
\email{reza699@ccs.neu.edu}

\author{Saman Jafari}
\affiliation{\institution{Northeastern University}}
\email{jafari1149@ccs.neu.edu}

\author{Sajjad Arshad}
\affiliation{\institution{Northeastern University}}
\email{arshad@ccs.neu.edu}

\author{William Robertson}
\affiliation{\institution{Northeastern University}}
\email{wkr@ccs.neu.edu}

\author{Engin Kirda}
\affiliation{\institution{Northeastern University}}
\email{ek@ccs.neu.edu}

\author{Hamed Okhravi}
\affiliation{\institution{MIT Lincoln Laboratory}}
\email{hamed.okhravi@ll.mit.edu}

\begin{abstract}

Control flow integrity (CFI) has received significant attention in the community
to combat control hijacking attacks in the presence of memory corruption
vulnerabilities. The challenges in creating a practical CFI has resulted in the
development of a new type of CFI based on runtime type checking (RTC). RTC-based
CFI has been implemented in a number of recent practical efforts such as
GRSecurity Reuse Attack Protector (RAP) and LLVM-CFI. While there has been a
number of previous efforts that studied the strengths and limitations of other
types of CFI techniques, little has been done to evaluate the RTC-based CFI. In
this work, we study the effectiveness of RTC from the security and practicality
aspects. From the security perspective, we observe that type collisions are
abundant in sufficiently large code bases but exploiting them to build a
functional attack is not straightforward. Then we show how an attacker can
successfully bypass RTC techniques using a variant of ROP attacks that respect
type checking (called \textit{\na}) and also built two proof-of-concept
exploits, one against Nginx web server and the other against Exim mail server.
We also discuss practical challenges of implementing RTC. Our findings suggest
that while RTC is more practical  for applying CFI to large code bases, its
policy is not strong enough when facing a motivated attacker.

\end{abstract}

\maketitle

\section{Introduction}

Memory corruption attacks continue to pose a major threat to computer systems.
Over the past decades, the sophistication of such attacks has risen from simple
code injection~\cite{smashing} to various forms of code-reuse attacks (a.k.a.
return-oriented programming --
ROP)~\cite{ccs2007rop,jitrop,blind,checkoway2010return} as a result of the
widespread adoption of defenses such as $W \oplus X$~\cite{DEP}.

Preventing memory corruption attacks in legacy, memory unsafe languages such as
C/C++ is challenging. Complete memory safety techniques that guarantee spatial
and temporal pointer safety often incur large runtime
overhead~\cite{softbound,cets}. As a result, lighter-weight defenses have been
proposed that enforce weaker policies, but incur lower performance overhead. One
class of such defenses randomizes or diversifies code at compile-time, 
load-time, or runtime~\cite{diversity} to create non-determinism for an attacker.
However, code randomization and diversification techniques are shown to be
vulnerable to various forms of direct~\cite{secrecy}, indirect~\cite{isomeron},
and side-channel~\cite{sidechannel} information leakage attacks. Even leakage-resilient 
variants of such defenses are shown to be vulnerable to code
inference~\cite{snow2016return} and indirect profiling attacks~\cite{aocr}.

A class of memory defenses that aims to provide a balance between security and
performance is Control Flow Integrity (CFI)~\cite{cfi_survey}. CFI aims to
prevent control hijacking memory corruption attacks by checking the control flow
transfers at runtime. While the policy enforced by CFI does not prevent 
non-control hijacking attacks (\eg data-only attacks~\cite{dop}), the relatively low
overhead incurred by CFI and its resilience to information leakage attacks make
it one of the desirable classes of defenses. CFI has even been called ``one of
the most promising ways to stop advanced code-reuse
attacks''~\cite{van2016tough}.

One of the distinguishing factors among various CFI techniques is how the control
flow graph (CFG) is generated. Three major classes of CFI defenses are: 1)
those that generate the CFG statically using points-to analysis
\cite{ccs2005cfi,wit,monitor,cfl,hypersafe},  2) those that generate the CFG
dynamically at runtime \cite{epcfi,bh,percfi}, and 3) those that  generate the
CFG based on type information \cite{rap,mocfi,kcfi,van2016tough,llvmcfi}.  We
call the third class Runtime Type Checking (RTC)-based CFI (or simply RTC in the
rest of this paper). Since points-to analysis is often very imprecise, difficult
to modularize, and hard when only the binary is available, many recent CFI
techniques are designed based on RTC \cite{rap,mocfi,kcfi,van2016tough,llvmcfi}.

In RTC, for forward edge protection, the type of function pointer and the target
are checked at each forward edge control transfer. A weaker subclass of RTC
techniques only checks the \textit{arity} (argument count) of forward edge
transfers, and not the precise type~\cite{van2016tough,usenixsec2014forwardcfi}.
For backward edge protection (\ie return address protection), the type of callee
is checked during the function epilogue. RAP~\cite{rap},
TypeArmor~\cite{van2016tough}, Kernel CFI (KCFI)~\cite{kcfi}, MCFI~\cite{mocfi},
IFCC~\cite{usenixsec2014forwardcfi}, and LLVM-CFI~\cite{llvmcfi} are some of the
examples of RTC techniques. While extensive work has been done on the
effectiveness of CFI based on points-to analysis
(e.g.,~\cite{ccs2015jujutsu,sp2014outofcontrol,usenixsec2014cfi-effectiveness,usenixsec2015cfb,stackdefiler}),
the strength of RTC has not been studied.

To the best of our knowledge, three implementations of RTC are available that
protect both forward and backward edges with type checking: KCFI~\cite{kcfi},
RAP~\cite{rap}, and MCFI~\cite{mocfi}. Other approaches such as
IFCC~\cite{usenixsec2014forwardcfi}, LLVM-CFI~\cite{llvmcfi}, and
TypeArmor~\cite{van2016tough} only protect the forward edge. While an
implementation of KCFI is not available, an open source version of RAP and LLVM-CFI 
are available. RAP and LLVM-CFI also provide the most stable implementations
of RTC as they are targeting production environments, and are not research
prototypes (RAP has even been applied to the Linux distribution, Subgraph OS
\cite{subgraphos}). In addition, RAP provides a more accurate CFG than LLVM-CFI
because it removes static functions from the target set, unlike LLVM-CFI.
Furthermore, neither RAP nor LLVM-CFI limit the target set to address-taken
functions. For these reasons, we focus our analysis on RAP in this paper.
In addition, we focus only on C programs because RAP C++ is not free\footnote{We
tried to obtain the commercial version of RAP, but unfortunately were not able
to do so because, based on our exchanges with the staff at GRSecurity, procuring
the commercial version actually requires contracting GRSecurity's security
service, and is not as simple as purchasing a software package for a fee.}.

In this paper, we provide the first study on the security and practicality of RTC.
From the security perspective, we illustrate that type collisions exist, and
are, in fact, very common in sufficiently large applications. While, at first
glance, it may appear that such collisions should be straightforward to exploit
(similar to attacks that leverage the imprecisions in points-to analysis-based
CFI), we show that practical exploits against RTC face a major challenge: it is
unlikely that a corruptible pointer has the exact collision with a desirable
function for an attack (\eg a system call). Indeed, we show that while
collisions are abundant, collisions with sensitive targets such as system calls
are, in fact, rare. We use a layered invocation method against RTC in which a
sensitive function is called indirectly through multiple layers of other calls
that eventually end in a call that has a collision with a corruptible function
pointer. In other words, a sensitive function is called from a corruptible
pointer through various layers of other functions. We call the sensitive
function, the function that collides with a corruptible pointer, and the other
layers Execution-Gadget (\gadgetthree), Collision-Gadget (\gadgetone), and
Linker-Gadget (\gadgettwo), respectively. Since this form of ROP attack respects
the type checking (thus bypassing RTC), we call it \textit{\name}.

In order to illustrate the practicality of \na, we build two proof-of-concept
exploits, one against Nginx and the other against Exim, that successfully hijack
control in the presence of RTC. Our exploits successfully bypass the open-source
version of RAP~\cite{rap}. Furthermore, we perform an analysis of exploitable
conditions in many popular applications and servers. Our findings indicate that
collisions are abundantly found in real-world applications, and that the gadgets
necessary for a {\na} attack (i.e., \gadgetone, \gadgettwo, and \gadgetthree)
are prevalent in popular servers. Our results suggest that, while RTC techniques
complicate successful attacks and are, in many cases, more practical than
points-to analysis-based CFI, on their own, they are not sufficient to prevent
control hijacking in the face of motivated attackers.

In summary, our contributions are as follows:

\begin{itemize}

\item We provide a first in-depth analysis of the effectiveness of RTC
techniques.

\item We illustrate a code reuse attack, \na, that can successfully bypass
precise RTC even in the absence of collisions with sensitive functions.

\item We build two proof-of-concept exploits against Nginx and Exim to show the
practicality of \na.

\item We analyze many popular applications and servers, and show that the
conditions necessary for a successful attack are abundantly found in the 
real-world.

\item We discuss the practical challenges of adopting RTC techniques in large
programs.

\end{itemize}

\section{Background and Problem Definition}

Lack of memory management in unsafe programming languages, such as C/C++, has
been introducing significant threats since 1988 when the first Internet worm
exploited a buffer overflow vulnerability in Fingerd~\cite{fingerd}. As a
result, there has been a continuous arms race between the development of attacks
and defenses.

Defenses in the memory corruption domain can be broadly categorized into
enforcement-based and randomization-based techniques. While randomization-based
techniques~\cite{diversity} are vulnerable to various forms of information
leakage (e.g.,~\cite{secrecy,blind,sidechannel}) attacks, enforcement-based
techniques~\cite{softbound,cets,ccs2005cfi} are resilient to such attacks. Full
memory safety techniques that enforce spatial~\cite{softbound} and
temporal~\cite{cets} safety on pointers are examples of enforcement-based
defenses. Lighter-weight defenses in the enforcement-based category impose more
relaxed policies on code execution at runtime, but provide better performance.
Control Flow Integrity (CFI)~\cite{cfi_survey} is an example of such a defense
that has received significant attention in the community over the past years,
and has even been deployed in real-world systems~\cite{cfguard,chromium}.

\subsection{Control Flow Integrity (CFI)}

CFI checks the indirect control transfers at runtime to prevent control
hijacking attacks~\cite{cfi_survey}. It checks forward-edges (\eg indirect jumps
and calls) and/or backward-edges (\eg function returns) to prevent the
corruption of indirect control transfers via memory bugs. While this policy is
weaker than full memory safety (for example, it does not prevent data-only
attacks~\cite{dop}), CFI aims to prevent the most pernicious types of memory
corruption attacks at a relatively low performance cost. 

CFI techniques can be categorized into three broad classes based on how they
generate their control flow graph (CFG). Perhaps the most widely studied class
of CFI is points-to analysis-based CFI as it was originally proposed by Abadi
\etal~\cite{ccs2005cfi}. In this technique, the CFG is constructed statically
using  points-to analysis~\cite{ccs2005cfi,wit,monitor,cfl,hypersafe}. Another
class of CFI techniques construct their CFG dynamically~\cite{epcfi,bh,percfi}. Dynamic CFIs need additional computations to identify and add new edges to the CFG during execution. In this paper, we do not study them.

The effectiveness of points-to analysis-based CFI crucially depends on the
ability to construct an accurate CFG. However, a sound and precise CFG is hard
to construct in the general case due to the undecidability of points-to
analysis~\cite{ramalingam1994undecidability,pldi2007dsa}. Coarse-grained
points-to analysis-based CFI techniques~\cite{ccfir,bincfi} tackle this problem
by grouping many branch targets together; however, such over-approximation is
shown to be too relaxed to prevent control hijacking
attacks~\cite{sp2014outofcontrol,usenixsec2014cfi-effectiveness}. Even
fine-grained points-to analysis-based CFI techniques are shown to be too
permissive to prevent all forms of control hijacking attacks, either because of
the imprecisions of static analysis~\cite{ccs2015jujutsu}, or because of the
versatility of functions like \texttt{printf()}~\cite{usenixsec2015cfb}. These
challenges along with the practical difficulties of generating and handling CFGs
in a modular way (e.g., dynamic loading) have motivated the development of a
third class of CFI based on Runtime Type Checking (RTC).

\subsection{Runtime Type Checking}

RTC matches the type signature of each indirect control transfer with its
target.  For forward-edge protection, RTC checks the type signature of a
function pointer and its target prior to each indirect control transfer. For
backward-edge protection, the type signature of the callee is stored before the
call site. Then this signature is checked during the epilogue of the callee to make
sure that such a type signature exists in the call site. In other words, RTC
limits the control flow of a program to respect the type signatures. It, thus,
implements a form of CFI that relies on types. The type matching can be enforced using
label-based annotations, in which labels are `hashes' of function signatures.
What constitutes the type signature requires careful considerations, and is
further discussed in Section~\ref{sec:rap}.

\subsection{Arity Checking}

Arity refers to the number of arguments of a function. Arity checking is a
strictly  weaker form of RTC, in which only the number of arguments of call
sites and  target functions are matched. In other words, arity checking is a
form of RTC in  which only one type exists for the argument. Hence, at runtime
only the \textit{number} of arguments are compared. Arity checking has been
implemented both at the source code (IFCC)~\cite{usenixsec2014forwardcfi} and
binary levels (TypeArmor)~\cite{van2016tough}. IFCC assigns each function to a
set based on its number of arguments, and only allows indirect calls to the set
with the correct number of arguments. TypeArmor, in addition to the number of
arguments, considers the return type as well. TypeArmor has been shown to prevent
advanced code reuse attacks such as COOP~\cite{COOP}.

\subsection{Reuse Attack Protector (RAP)}
\label{sec:rap}

\begin{figure}
  \center
  \centerline{\includegraphics[width=\linewidth]{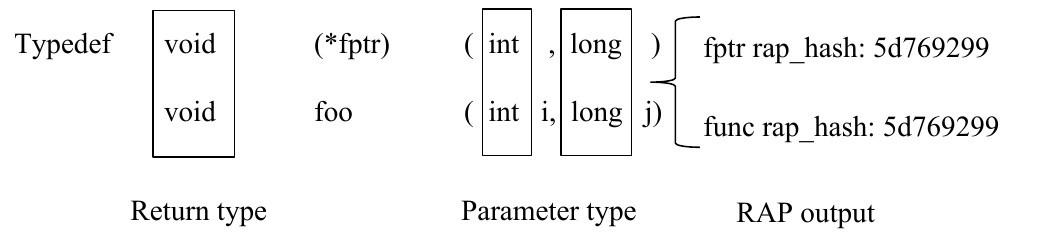}}
  \vspace{-.1in}
  \caption{Type parts in RAP.}
  \label{fig:typeRAP}
\end{figure}

RAP is a code-reuse attack protection that implements RTC.
Figure~\ref{fig:typeRAP} shows an example of type parts for a function pointer
and a function that are used in RAP. In this case, which shows a function and a
function pointer, the return type (\texttt{void}) and the argument types
(\texttt{int} and \texttt{long}) are parts of the type signature, while the
names are not. RAP generates two hashes for each function. One of them is used
when the function is the target of a function pointer. The other one is used
during backward-edge checking.

RAP calculates a hash for each function pointer and instruments the call site,
as shown in the line 4 of Figure~\ref{fig:rapforward}. This line compares the
expected hash with the function's type hash that is located before the actual
function's memory address. If the hashes do not match, the execution jumps to an
error; otherwise, the indirect call is taken as shown in line 6. Similarly, for
backward-edges, the hash value that is located before the call site and the one in
epilogue are compared and if they match, the program continues to return as
shown in Figure~\ref{fig:rapbackward}. Previous researches have shown that
backward-edge protection without enough sensitivity leads to the flexibility to
return to different call sites~\cite{sp2014outofcontrol,usenixsec2015cfb}. To
address this issue, RAP encrypts the return address with a key which is stored
in a register. This technique makes a context sensitive version of RAP. However,
this feature is not available in open source version of RAP. Note that RTC
without a shadow stack or an equivalent technique such as encrypted return address
is vulnerable to  backward-edge attacks that have been studied comprehensively
by other researchers~\cite{sp2014outofcontrol,usenixsec2015cfb}. Hence, for the
rest of this paper, we only focus on forward-edges.

\begin{figure}
  \begin{lstlisting}
  dq 0x11223344
  func:
  ...
  cmpq $0x11223344,-8(%rax)
  jne .error
  call *%rax  
  \end{lstlisting}
  \vspace{-.2in}
  \caption{Forward-edge checking in RAP.}
  \label{fig:rapforward}
\end{figure}

\begin{figure}
  \begin{lstlisting}
  jmp label
  dq 0xffffffffaabbccdd
  label:
  call func
  
  func()
  .
  .
  mov %(rsp),%rcx
  cmpq $0xaabbccdd,2(%rcx)
  jne .error
  retn
  \end{lstlisting}
  \vspace{-.2in}
  \caption{Backward-edge checking in RAP.}
  \label{fig:rapbackward}
\end{figure}

\subsection {Type Collisions}

\begin{figure}
  \footnotesize
  \begin{lstlisting}
typedef void (*FunctionPointer)(void);

int flag = 0;
char *cmd;

void valid_target1(void){
    printf("Valid Target 1\n");
}

void valid_target2(void){
    printf("Valid Target 2\n");
}

int final_target(char *cmd){
    system(cmd);
}

int linker_func(void){
    if (flag ==1)
      final_target(cmd);
}

void invalid_target(void){
    linker_func();
}

void vulnerable(char * input){
    FunctionPointer corruptible_fptr;
    char buf[20];

    if (strcmp(input, "1") == 0)
        corruptible_fptr = &valid_target1;
    else
        corruptible_fptr = &valid_target2;

    printf(input);
    strcpy(buf, input);

    corruptible_fptr();
}
  \end{lstlisting}
\vspace{-.2in}
  \caption{A sample vulnerable program.}
  \vspace{-.2in}
  \label{fig:vulnerabelcode}
\end{figure}

An astute reader might suspect that type collisions should exist in sufficiently
large code bases. This is indeed correct; however, we show that type collisions
do not immediately indicate the feasibility of a practical attack against RTC.
To clarify and explain, let us start with an example.

\begin{figure}
  \center
  \centerline{\includegraphics[width=\linewidth]{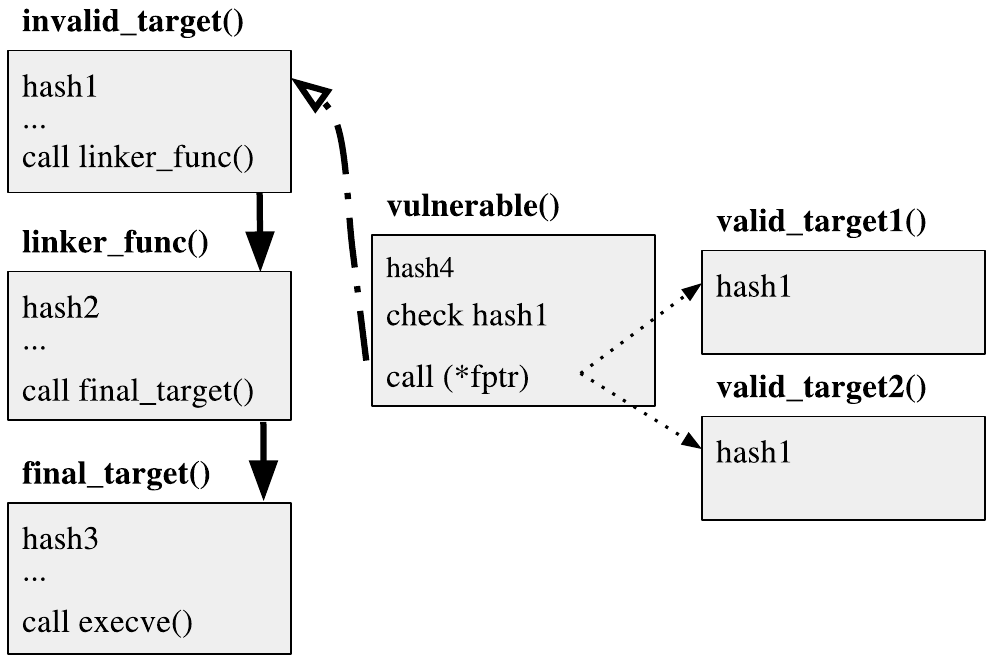}}

  \caption{Control flow graph of the sample vulnerable program. This figure
  illustrates how type collision leads to over-approximation.}

  \label{fig:vulncfg}
\end{figure}

Figure~\ref{fig:vulnerabelcode} shows a sample source code with intentional
vulnerabilities in lines 36 and 37 for leaking and overwriting. This code is
compiled by the RAP plugin that protects the code with RTC.
Figure~\ref{fig:vulncfg} shows its CFG. Although type collisions exist in this
code (between \emph{invalid\_target()} and \emph{corruptible\_fptr}), according
to RTC, it is not allowable to call functions such as \emph{linker\_func()} or
\emph{final\_target()} (which may be interesting targets for an attacker because
of their ability to spawn a malicious shell) with the \emph{corruptible\_fptr}
function pointer. Such a call is prohibited because the type of the functions
and the \emph{corruptible\_fptr} function pointer are different. As a result, while many
type collisions might exist in large code bases, their usefulness for a
practical attack is questionable. There may not exist any sensitive target that
has type collision with a corruptible function pointer. Indeed in our analysis
of real-world code bases, we rarely found a package in which a sensitive
function (\eg a system call) collides with a corruptible function pointer. Thus,
the questions regarding the effectiveness of RTC are not trivial to answer.

However, we make an observation in this sample source code that provides an
insight into how an attack can be built. We observe that while the sensitive
functions cannot directly be called by the 
\emph{corruptible\_fptr} function pointer, it is still possible to invoke these functions through
other functions. In this case, the \emph{invalid\_target()} function has the
same type as the \emph{corruptible\_fptr} function pointer, so it is feasible to call this function,
and as it can be seen, there is a path from this function to the
\emph{final\_target()} function. At least in theory, it looks like that it should be
feasible to reach the \emph{final\_target()} function indirectly from the
\emph{corruptible\_fptr} function pointer. To do so; however, one must consider the constraints
in the execution path such as the \textit{if} condition in line 19. In this
case, the constraint is satisfiable because the \textit{if} condition checks a
global variable which can be overwritten by the attacker.

\subsection{Research Questions}

The above example and the observations made about possible collisions and
indirect invocations of sensitive functions raise the following research
questions:

\begin{itemize}

\item Can RTC be practically bypassed using type collisions?

\item Are there enough intermediate functions with satisfiable constraints in
real-world applications that allow an attacker to hijack control to a sensitive
function (\eg system calls) in the presence of RTC?

\item How prevalent are these constructs in real-world applications?

\end{itemize}

In the rest of the paper, we provide answers to these questions in order to
evaluate the effectiveness of the RTC from the security perspective in
Sections~\ref{sec:attack}, \ref{sec:exploit}, and \ref{sec:eval}. We discuss the
practicality considerations in Section \ref{sec:disc}.

\section{Attack Overview}
\label{sec:attack}

In order to evaluate the effectiveness of the RTC, we show how an attacker can
exploit type collisions to build real attacks. Although RTC significantly
reduces the number of valid targets, we show that the scope and prevalence of
collisions make it possible to attack many real-world applications.

\subsection {Threat Model}

Our threat model is inline with the larger body of literature in the area of
memory corruption. Since data execution prevention (a.k.a. DEP or W $\oplus$ X)
and address space layout randomization (ASLR) are widely deployed in modern
operating systems, we assume that they are enabled on the target. Moreover, we
assume that RAP is also deployed on the target. Any control hijacking attempt
that violates type checking is properly detected and stopped. Moreover, since we
are interested in studying the effectiveness of the RTC paradigm as a whole, and
not the quality of any particular implementation, we consider
\textit{implementation} bugs or flaws out-of-scope for our attacks. In other
word, we do not target any implementation weakness.

On the attacker side, we assume the target application contains one strong or
multiple limited memory corruption vulnerabilities that allow an attacker to
write arbitrary values to arbitrary memory locations including stack and heap
and also leak some information by arbitrary read primitive. A strong
vulnerability similar to CVE-2017-7184 (exploited in Pwn2Own 2017),
CVE-2017-0143 (exploited in WannaCry), CVE-2016-4117, or CVE-2015-0057
simultaneously provides arbitrary read and write primitives, and can be
exploited multiple times to overwrite multiple values in 
memory~\cite{phrack-multioverwrite}. A more limited set of vulnerabilities can alternatively be
used to first read memory, and then write to it maliciously. For instance, array out-of-bound access (CVE-2018-5008), non-terminated 
strings (CVE-2017-7790) and format string can be used for arbitrary read primitive.
Other vulnerabilities such as double free (CVE-2018-4990), type confusion (CVE-2015-1641) and format string can
be used for arbitrary write primitive. The long history of memory corruption
vulnerabilities has demonstrated that assuming their existence even in the most
tested code bases (\eg Linux kernel and Windows Services) is a reasonable and
often valid proposition. The assumptions in this work are thus realistic and 
inline with the defenses~\cite{readactor,readactor2,aslrg,tasr,van2016tough} and
attacks~\cite{COOP,blind,aocr,jitrop,isomeron} presented in the literature.

\subsection {Attack Preliminaries}

In order to perform a successful \name attack against RTC, an adversary needs to
follow some steps to execute an arbitrary code. First, during the offline
preparation phase, the attacker searches for an interesting function that allows
arbitrary execution, for example \emph{execve()} or any other function that has
an equivalent behavior. To do so, the arguments of an intended function must be
controllable by the attacker. For example, if the arguments reside on heap or
global memory, they can be overwritten at anytime by an attacker. Next, the
attacker needs to change a function pointer to hijack control. It is imperative
to note that the type of the function pointer and the function should be the
same; otherwise, RTC will prevent the control transfer. This is what
distinguishes a \na~  attack from a traditional ROP attack; \ie a \na~attack is
a special form of ROP in which the hijacked control respects the type checking.

Although this attack looks powerful in that it allows arbitrary code execution,
as mentioned earlier, it is very unlikely to find such a type collision in
real-world applications. This is due to the fact that the attacker needs to have
access to a corruptible pointer that has the same type as the sensitive function
of interest for hijacking (in this case \emph{execve()}). In order to address
this challenge, we use a technique based on layered invocations and type
collisions, which gives the attacker the opportunity to call functions even with
differing types from a corruptible function pointer. In a nutshell, a target
function of interest (\ie a sensitive function) is invoked indirectly through
other intermediary functions in such a way that: 1) the first invoked function
has the same type signature as the corrupted function pointer, 2) each function
in the chain contains a valid invocation of the next function in its code, and
3) the last function in the chain is the target function of interest.

To facilitate the description and analysis of \na, we introduce three types of
gadgets that make it possible. We call the first function that has a type
collision with a corruptible function pointer a Collision gadget (\gadgetone),
the target function that the attacker intends to invoke maliciously an Execution
gadget (\gadgetthree), and the intermediary functions that are called in a
nested fashion (from the collision gadget to the execution gadget) Linker
gadgets (\gadgettwo).

Figure~\ref{fig:gadgetgeneration} illustrates the high-level view of the
location of the gadgets. Each node is a representation of a gadget. The attacker
first corrupts a function pointer to redirect control to the \gadgetone, which,
by definition, respects the type checking. The \gadgetone then calls other
{\gadgettwo}s that, in turn, eventually call the \gadgetthree.

\subsection{Finding Gadgets}

In this section, we describe the method to find and chain proper gadgets to
perform a \na~attack. In our terminology, we define the whole function as a
gadget. Finding gadgets starts with a corruptible function pointer. For example,
any function pointer on the stack or heap adjacent to an overflow-able buffer
(\eg CVE-2016-9679), or those that have known or leakable addresses (\eg
CVE-2017-7219) are potential targets for corruption. The ideal scenario is to
point this pointer to the final target, but due to RTC, this is not possible in
most cases, so instead, the control is transferred to a \gadgetone. Any function
that has the same type signature as the corruptible function pointer is a
candidate for a \gadgetone. The next step is to find an appropriate
\gadgetthree. For \gadgetthree, we chose to focus on functions that spawn a
shell (\eg \texttt{execve} system call or their wrappers) since that provides an
attacker with a wide range of malicious capabilities sought after in the
payloads. However, an attacker may choose more limited functions (\eg one that
disables DEP) if those are sufficient for the ultimate purpose of the attack.
After selecting the \gadgetone and \gadgetthree, we need to find proper
{\gadgettwo}s to chain them together. With the help of the program's call graph,
both direct and indirect calls, {\gadgettwo}s can be found by traversing all the
candidate paths starting from a {\gadgetone} and ending in an {\gadgetthree}.

We have developed a semi-automated tool that receives as input the set of all
{\gadgetone}s and {\gadgetthree}s as well as the program's call graph, and finds
all candidate {\gadgettwo}s in a given code base. The tool uses RAP's output
during compilation which is in verbose mode. The call sites and target sets are
identifiable in this output. The simplified algorithm inside our tool for
finding all candidate paths is listed in
Algorithm~\ref{snippet:candidate_paths}. In the algorithm:

\begin{enumerate}

\item $G$ is the adjacency matrix of the program's direct and indirect call
graph

\item $N$ is the set of all nodes (functions) in the graph

\item $CG$ is the set of all {\gadgetone}s

\item $EG$ is the set of all {\gadgetthree}s

\item $CP$ is the set of all candidate paths

\item $cg$ is the {\gadgetone}

\item $eg$ is the {\gadgetthree}

\item $V$ is the set of already visited nodes (functions) in the graph in
order to prevent loops through recursive path discovery algorithm
(\textsc{DiscoverPaths})

\item $P$ is an ordered list of functions in the call chain

\end{enumerate}

\newcommand{\diffblock}[1]{#1}

\algnewcommand\True{\textbf{true}\space}
\algnewcommand\False{\textbf{false}\space}
\algnewcommand\Andd{\textbf{and}\space}
\algnewcommand\Break{\textbf{break}\space}

\diffblock{
\begin{algorithm}
\caption{Finding candidate paths.}
\label{snippet:candidate_paths}
{
\footnotesize
\begin{algorithmic}
\Function{FindCandidatePaths}{$G, N, CG, EG$}
  \State $CP\gets \emptyset$
  \\
  \For{$cg \in CG$}
    \For{$eg \in EG$}
      \State \Call{DiscoverPaths}{$CP, cg, eg, \emptyset, \emptyset$}
    \EndFor
  \EndFor
  \\
  \State \Return $CP$
\EndFunction
\\
\Function{DiscoverPaths}{$CP, cg, eg, P, V$}
  \State $P\gets P\cup \{cg\}$, $V\gets V\cup \{cg\}$
  \\
  \If{$cg == eg$}
    \State $CP\gets CP\cup \{P\}$
  \Else
    \For{$g \in N$}
      \If{$g \not \in V$ \Andd $G[cg][g] = 1$}
        \State \Call{DiscoverPaths}{$CP, g, eg, P, V$}
      \EndIf
    \EndFor
  \EndIf
  \\
  \State $P\gets P - \{cg\}$, $V\gets V - \{cg\}$
\EndFunction
\end{algorithmic}
}
\end{algorithm}
}

As it can be seen in Algorithm~\ref{snippet:candidate_paths}, the
\textsc{FindCandidatePaths} function iterates over all {\gadgetone}s and
{\gadgetthree}s, and by using \textsc{DiscoverPaths}, it finds all the possible
paths (\textit{candidate} paths) between every combination of {\gadgetone} and
{\gadgetthree}. Finding proper {\gadgettwo}s is more challenging though. We
define an ideal \gadgettwo as one that satisfies these conditions:

\begin{enumerate}

\item The constraints inside the gadgets are controllable from outside of it.
For example, the constraints are based on global variables that could be
maliciously modified from outside of the gadget.

\item There is no constraint inside of the {\gadgettwo}s that affects the flow
to the \gadgetthree.

\end{enumerate}

\begin{figure}
    \centerline{\includegraphics[width=\columnwidth]{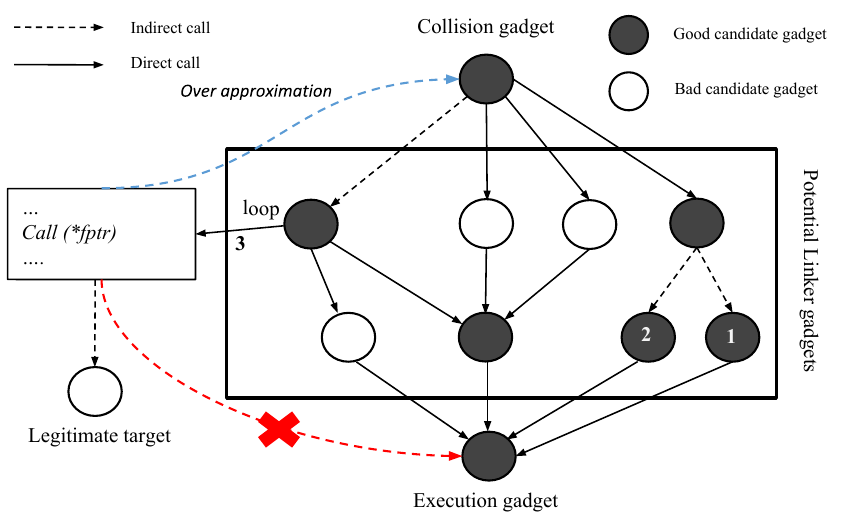}}

    \caption{The location of gadgets in the CFG of the program. Due to the type
    mismatching, the \texttt{fptr} is not able to point to the execution gadget
    directly; hence it utilizes over approximation edge.}

    \label{fig:gadgetgeneration}
\end{figure}

As it can be seen in Figure~\ref{fig:gadgetgeneration}, there might be multiple
paths from a \gadgetone to an \gadgetthree. Some paths are useful while others
are not. Empty nodes depict {\gadgettwo}s that do not have at least one of the
two attributes mentioned above, so they are eliminated. The shaded nodes depict
good candidates to reach the target. There are also some cases where there is a
pointer that can be used to switch between different paths (in this example,
nodes 1 and 2). This provides more flexibility to the attacker.

We also note that it is possible to create a loop using these gadgets. Creating
a loop gives the opportunity to the attacker to trigger a vulnerability multiple
times. This is depicted in Figure~\ref{fig:gadgetgeneration} by edge \#3.

\subsection{Constraint Solving}

As we consider the whole function as a gadget, there might be some conditions in
{\gadgettwo}s which influence the control flow of the program. If an \gadgettwo
contains a condition that prevents reaching the intended \gadgetthree, then it
needs to be resolved. For instance, if there is an \emph{if} condition in the
\gadgettwo that returns from the function, and halts the flow to the invocation
point of the \gadgetthree, or if there is a null data structure that is needed
to be filled with proper data before the \gadgettwo can be executed, additional
constraints need to be satisfied. Therefore, the list of constraints between a
{\gadgetone} and an {\gadgetthree} is every point in the program that changes
the flow between these two nodes. To see if the constraints before that point
are satisfiable, the following conditions should be in place:

\begin{enumerate}

\item The constraint can be solved by overwriting data in memory.

\item The data should be also accessible globally in memory outside the
function.

\end{enumerate}

The first condition ensures that the constraint can actually be controlled by a
malicious memory overwrite, while the second condition ensures that the window
of opportunity for overwriting memory is not too short. For example, while local
variables may become corruptible during limited windows of time, their sometimes
short longevity makes them inappropriate targets for constraint solving. In the
last step, to get the concrete values for the constraints, depending on the
number of constraints and complexity of them, either manual solving or symbolic
execution are applicable. For our PoC exploits, we solved the constraints
manually.

\section{Proof-of-Concept Exploits}
\label{sec:exploit}

As a proof-of-concept, we show how the Nginx web server and the Exim mail server
protected by RTC can be exploited by \na~attacks. In both cases, we manage to
achieve arbitrary execution while the server is protected by open source version
of RAP. We show how an attacker is able to redirect control to a function that
calls a function from the \emph{exec} family. This enables us to execute
arbitrary commands via the spawned shell, which is a powerful attack. To do so,
we compiled Nginx-1.10.1 and Exim-4.89 with the RAP GCC plugin. As we assumed in our threat model that we have arbitrary read and write access to the memory, \texttt{gdb} is used to mimic these primitives.

\subsection {Nginx Exploit}

\begin{figure}
  \center
  \centerline{\includegraphics[width=\linewidth,scale = 0.4]{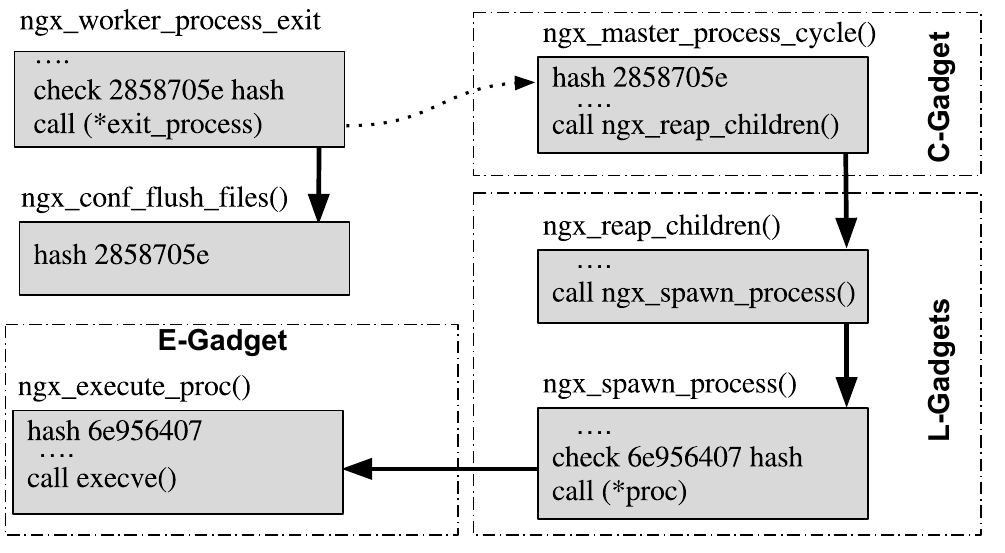}}
  \vspace{-.1in}
  \caption{The flow of the exploit in Nginx and the corresponding gadgets.}
  \vspace{-.1in}
  \label{fig:nginxgadgets}
\end{figure}

In the best case scenario, an attacker needs at least two types of gadgets,
\gadgetone and \gadgetthree. However, our analysis of Nginx using our tool
indicates that there is no direct collision between a potentially corruptible
function pointer and a sensitive function (\eg a system call). We thus need to
find enough {\gadgettwo}s to link corruptible function pointers with a sensitive
function such as \emph{execve()}. Searching through Nginx's code using the
algorithms described in the previous section, we can indeed find enough
{\gadgettwo}s with satisfiable constraints.

The overview of our Nginx attack is illustrated in
Figure~\ref{fig:nginxgadgets}. The attack starts by corrupting the
\emph{exit\_process} function pointer. Both the \emph{exit\_process} function
pointer and the \emph{ngx\_master\_process\_cycle()} function have the same type
signature, as it can be seen in Figure ~\ref{fig:ngxMasterProcessCycle}. Therefore, \emph{ngx\_master\_process\_cycle()} function can be used as our
\gadgetone. Due to the matching type signatures, such a malicious redirection is
not detected by RTC. Even though \emph{ngx\_conf\_flush\_files()} function is
the the valid target of the \emph{exit\_process} function pointer, due to the
type collision, we can change it to point to the
\emph{ngx\_master\_process\_cycle()} function instead. Our analysis indicates
that the \emph{ngx\_execute\_proc()} function, which calls \emph{execve()}
directly, can be a great candidate for our \gadgetthree, so we use it in our
attack. Figure~\ref{fig:ngx_execute_proc} shows that the parameters of the \emph{execve()} function are also controllable in a global data structure. Now that we have our \gadgetone and \gadgetthree, we have to chain them
with enough {\gadgettwo}s. Our analysis based on the call graph of Nginx
indicates that two good candidates for {\gadgettwo}s are the
\emph{ngx\_reap\_children()} and the \emph{ngx\_spawn\_process()} functions. Figure~\ref{fig:ngx_spawn_process} shows there is the \emph{proc} function pointer in our \gadgettwo which provides more flexibility for the attacker to change the flow.

\begin{figure}
  \begin{lstlisting}
  // Type definition of exit_process pointer
  void (*exit_process)(ngx_cycle_t *cycle)
  // Type definition of ngx_master_process_cycle
  void ngx_master_process_cycle(ngx_cycle_t *cycle)

  void ngx_master_process_cycle(ngx_cycle_t *cycle) {
    ...
    // This function helps to create a loop. It calls (*exit_process) in the following
    ngx_start_worker_processes(cycle, ccf->worker_processes, NGX_PROCESS_RESPAWN);
    ...
    // By setting this condition to true, the attacker can reach to the next gadget which is ngx_reap_children()
    if (ngx_reap) {
        ngx_reap = 0;
        ngx_log_debug0(NGX_LOG_DEBUG_EVENT, cycle->log, 0, "reap children");
        live = ngx_reap_children(cycle);}

  \end{lstlisting}
  \vspace{-1em}
  \caption{\texttt{ngx\_master\_process\_cycle} is a {\gadgetone}}
  \label{fig:ngxMasterProcessCycle}
\vspace{-0.5em}
\end{figure}

\paragraph{Constraint Solving}

There are some constraints in our {\gadgetone} and {\gadgettwo}s that should be
solved. For example, there is an \textit{if} condition in the
\emph{ngx\_master\_process\_cycle()} function on \textit{ngx\_reap} variable
that might prevent execution from reaching \emph{ngx\_reap\_children()}.
Moreover, there are multiple \textit{if} conditions in the
\emph{ngx\_reap\_children()} function, our first {\gadgettwo}, that have to be
satisfied to call the \emph{ngx\_spawn\_process()} function, our second
{\gadgettwo}. Our second {\gadgettwo} (\emph{ngx\_spawn\_process}) always calls
our {\gadgetthree} without any interruption, so in this case, there is no
constraint to be satisfied.

In order to satisfy the constraints listed above, we observe that they are all
controllable by overwriting global variables or heap objects. Consequently, in
our exploit, we first use our arbitrary write vulnerability to set the value of
\emph{ngx\_reap} that resides in global memory to 1. This allows our \gadgetone
(\emph{ngx\_master\_process\_cycle()}) to call our first \gadgettwo
(\emph{ngx\_reap\_children()}). We then use the same vulnerability to write the
desired values to \textit{ngx\_processes[i].respawn},
\textit{ngx\_processes[i].exiting}, \textit{ngx\_terminate}, and
\textit{ngx\_quit} on the heap that allows our first \gadgettwo
(\emph{ngx\_reap\_children()}) to call our second \gadgettwo. Then we overwrite
the function pointer (\emph{exit\_process}) to point to our \gadgetone. After
this overwrite, the execution passes from \gadgetone to the {\gadgettwo}s to the
\gadgetthree, at which point a malicious shell is spawned under our control.

\begin{figure}
  \begin{lstlisting}
  ngx_execute_proc(ngx_cycle_t *cycle, void *data)
  {
      ngx_exec_ctx_t  *ctx = data;
      if (execve(ctx->path, ctx->argv, ctx->envp) == -1) {
          ngx_log_error(
          NGX_LOG_ALERT, 
          cycle->log, 
          ngx_errno,
          "execve() failed while executing %s \"%s\"",
          ctx->name, ctx->path);
      }
      exit(1);
  }
  \end{lstlisting}
  \vspace{-1em}
  \caption{\texttt{ngx\_execute\_proc} is an {\gadgetthree}. The parameters of \texttt{execve} function are controllable globally.}
  \label{fig:ngx_execute_proc}
 \vspace{-0.5em}
\end{figure}

\begin{figure}
  \begin{lstlisting}
  ngx_spawn_process(
    ngx_cycle_t *cycle, 
    ngx_spawn_proc_pt proc, 
    void *data,
    char *name, 
    ngx_int_t respawn) {
  ...
    proc(cycle, data); // The proc function pointer invokes ngx_execute_proc function in this case
  }
  \end{lstlisting}
   \vspace{-1em}
  \caption{\texttt{ngx\_spawn\_process} is an {\gadgettwo}}
  \label{fig:ngx_spawn_process}
 \vspace{-0.5em}
\end{figure}

\subsection {Exim Exploit}

\begin{figure}
    \center
    \centerline{\includegraphics[width=\linewidth,scale = 0.4]{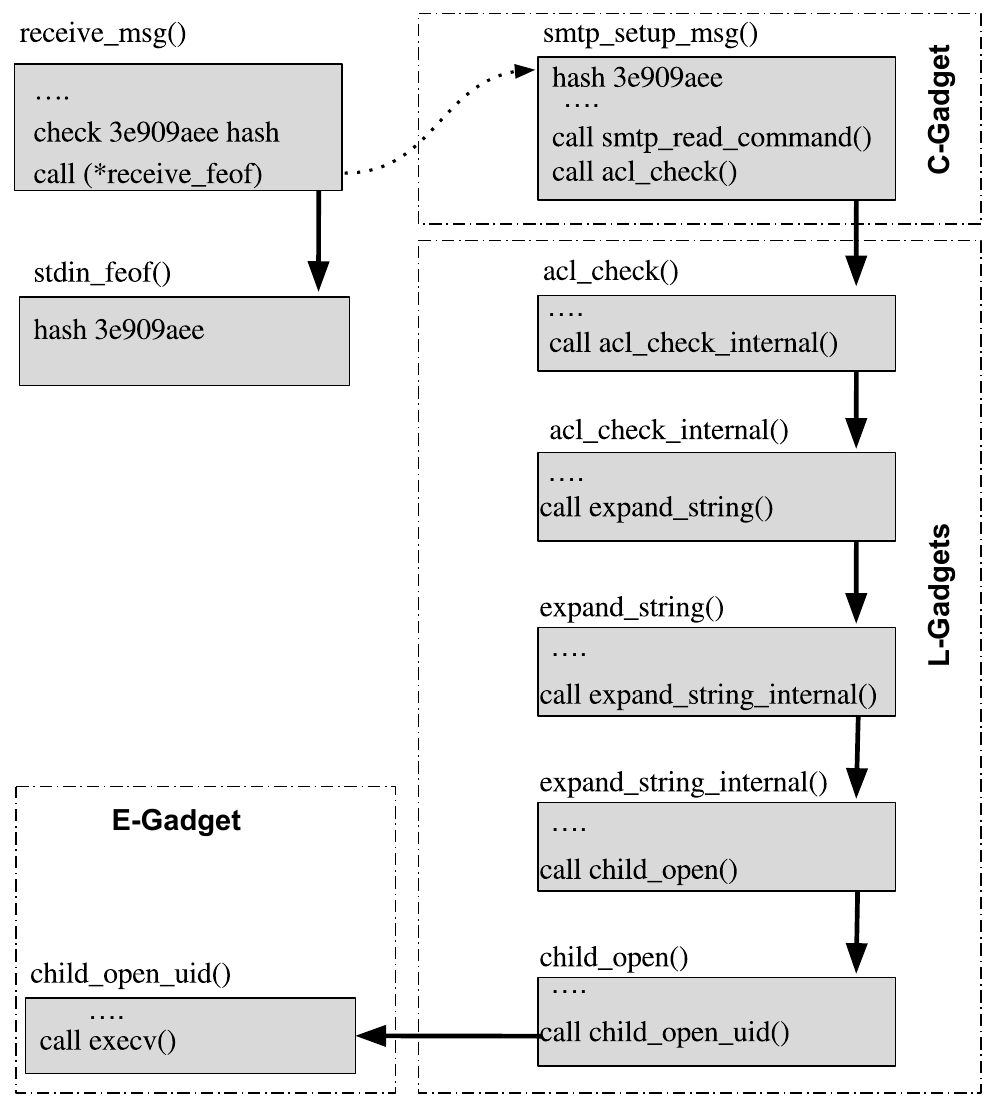}}
    \vspace{-.1in}
    \caption{The flow of the exploit in Exim and the corresponding gadgets.}
    \vspace{-.1in}
    \label{fig:eximgadgets}
\end{figure}

For the sake of brevity and due to its similarity to Nginx exploit, we skip the
details of the steps necessary to build Exim exploit. We find that Exim's source
code is indeed large enough that it contains proper gadgets, and a large number
of corruptible memory read and write operations. A previous example of such a
corruption happened with CVE-2016-9963 that allowed a remote attacker to read
the private DKIM signing key, and write it to a log file. In our exploit, we
again focus on spawning a malicious shell because of its generality and strength
as a remote attack.

Like our Nginx exploit, we use our gadget finder tool to find the proper gadgets
and paths in the program that can be used to reach an \gadgetthree, in Exim's
case, the \emph{child\_open\_uid()} function. We maliciously overwrite the
\emph{receive\_feof} function pointer inside the \emph{receive\_msg()} function
and redirect execution to the \emph{smtp\_setup\_msg()} function, which serves
as our \gadgetone. The functions \emph{acl\_check()},
\emph{acl\_check\_internal()}, \emph{expand\_string()},
\emph{expand\_string\_internal}, and \emph{child\_open()} serve as our
{\gadgettwo}s with easily satisfiable constraints. Note that while the chain of
gadgets is much longer in the case of Exim, the exploit is indeed simpler
because there are fewer constraints in the {\gadgettwo}s. The flow of the
exploit and the position of gadgets are shown in Figure~\ref{fig:eximgadgets}.
In order to bypass some constraints of the {\gadgettwo}s, we used
\texttt{\$\{run\{/bin/bash\}\}\$\textbackslash\textbackslash} as the payload.
This complex payload helps us execute the shell command and solve the
constraints.

\subsection{Summary}

These two exploits indicate that although RTC does complicate a successful
exploit and places a number of limitations on how such an exploit can be built,
motivated attackers may still find feasible and realistic opportunities to build
damaging \na~attacks using type collisions.

\section{Evaluation}
\label{sec:eval}
\begin{figure*}[t]
    \centering
    {\sffamily
    \begin{subfigure}[t]{0.45\linewidth}
        \includegraphics[width=\textwidth]{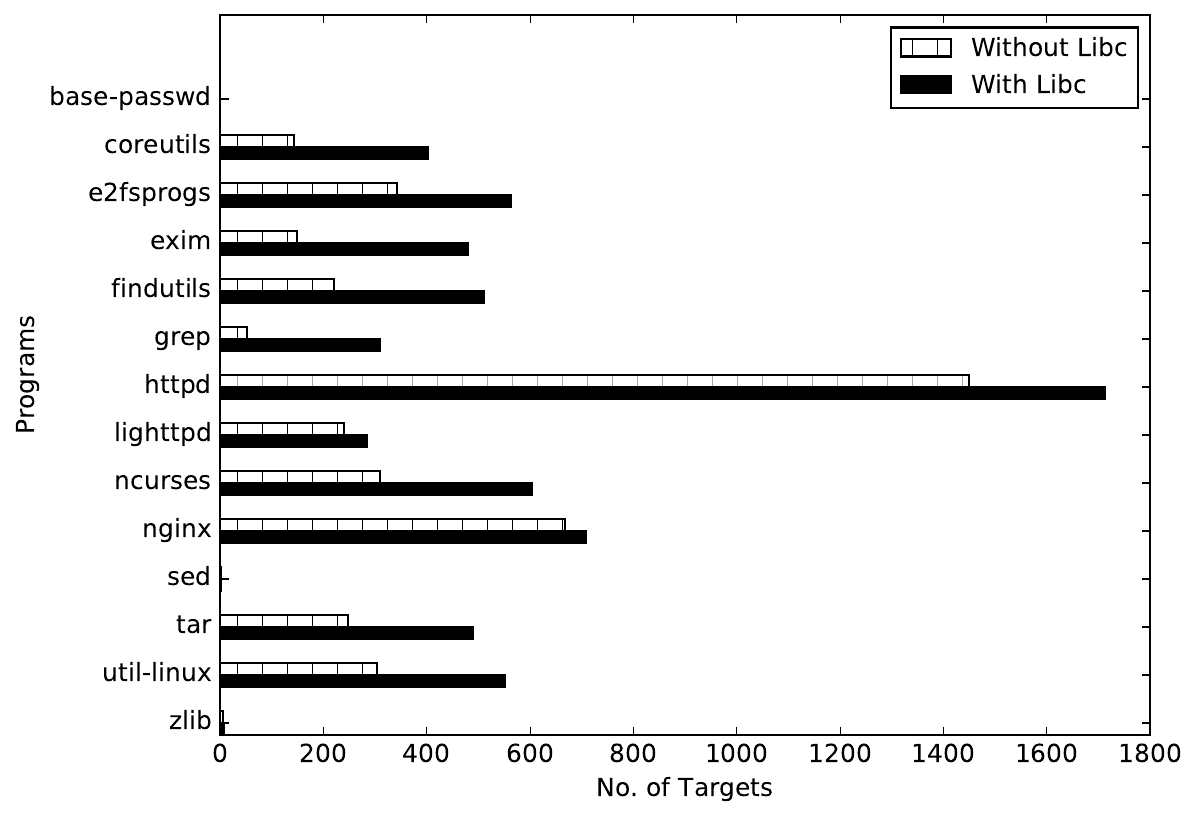}
        \caption{Targets}
        \label{fig:libc:targets}
    \end{subfigure}
    \hfill
    \begin{subfigure}[t]{0.45\linewidth}
        \includegraphics[width=\textwidth]{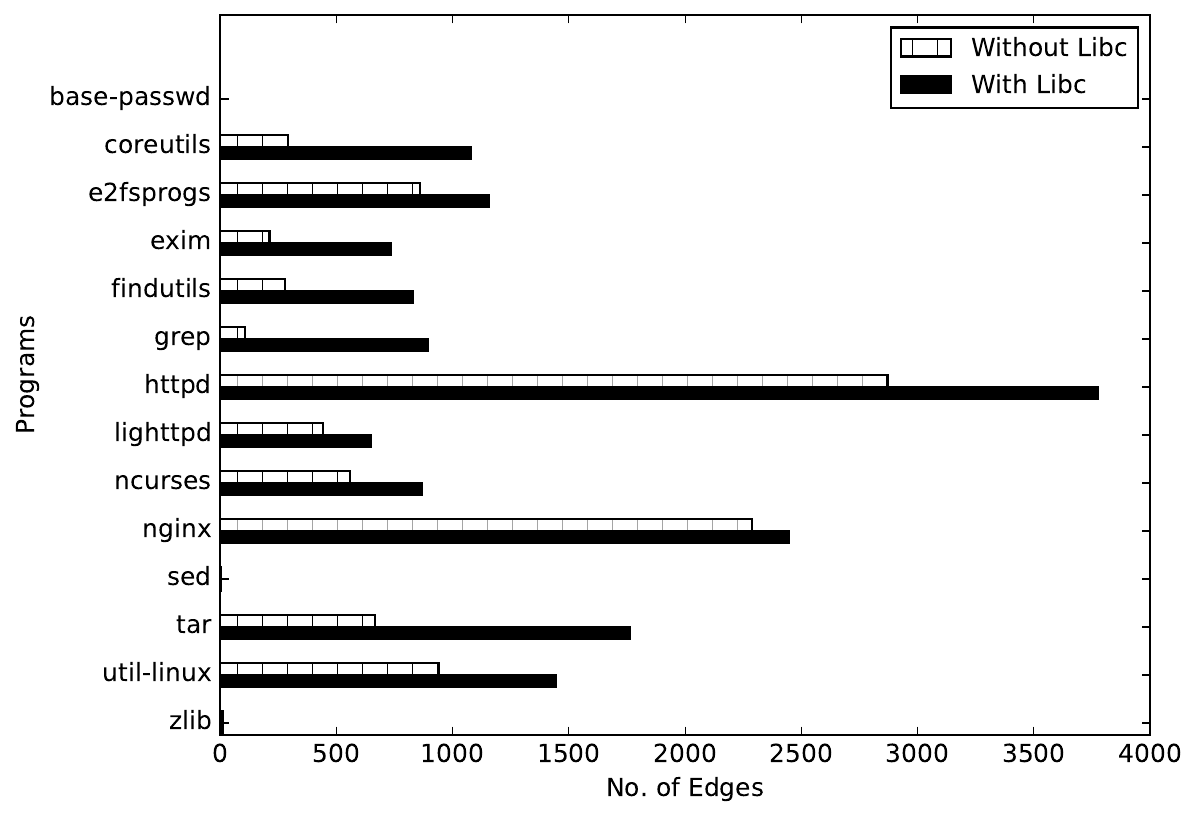}
        \caption{Edges}
        \label{fig:libc:edges}
    \end{subfigure}
    \vspace{-0.4em}
    \caption{Increase in the number of targets and edges when linked with \texttt{glibc}.}
    \vspace{-1em}
    \label{fig:libc}
    }
\end{figure*}

In this section, we evaluate the prevalence of collisions and gadgets necessary
for launching a \na~attack. For our evaluation, we chose the top 10 C packages
in the Ubuntu repository~\cite{UbuntuPopularityContest} as well as three widely
used web servers (Httpd, Nginx, and Lighttpd) and a mail server (Exim). The list
of the applications are shown in Table~\ref{tab:over_approximation}. For each of
these applications, we compiled it with the RAP GCC plugin. We then used our
gadget finder tool that parses the RAP's output and generates a JSON file
containing the function pointers as well as their call sites in the program. In
addition, our tool extracts the target functions for each of these call sites.
In all of our evaluation, the default modules of the analyzed applications were
used. In addition, the open source version of RAP was deployed with default
options and compilation flags.

\subsection{Type Collisions}

\begin{table*}[t]
  \caption{Over-approximantion of target functions and indirect calls because of type collisions.}
  \label{tab:over_approximation}
\small
  \centering
  \begin{tabular}{llrrrrrrrrrrrrr}

  \toprule

  \makecell{\multirow{2}{*}{\textbf{App}}} &
  \makecell{\multirow{2}{*}{\textbf{Version}}} &
  \multirow{2}{*}{\textbf{\makecell{Function\\Pointer}}} &
  \multirow{2}{*}{\textbf{\makecell{Call\\Sites}}} &
  \makecell{\multirow{2}{*}{\textbf{Functions}}} &
  \multirow{2}{*}{\textbf{\makecell{Functions\\w/ Hash}}} &
  \multicolumn{2}{c}{\textbf{Function Targets}} &
  \multicolumn{2}{c}{\textbf{Indirect Calls}}
  \\

  \cmidrule[0.5pt](lr){7-8}
  \cmidrule[0.5pt](lr){9-10}

  &
  &
  &
  &
  &
  & \textbf{\makecell{All}}
  & \textbf{\makecell{Invalid}}
  & \textbf{\makecell{All}}
  & \textbf{\makecell{Invalid}}
  \\

  \midrule

base-passwd & 3.5.39 & 6 & 6 & 45 & 45 (100.0\%) & 0 & 0 (0.0\%) & 0 & 0 (0.0\%) \\
coreutils & 8.2 & 42 & 80 & 1,789 & 682 (38.1\%) & 116 & 43 (37.1\%) & 416 & 110 (26.4\%) \\
e2fsprogs & 1.42.13 & 97 & 264 & 1,964 & 1,243 (63.3\%) & 251 & 176 (70.1\%) & 1,383 & 400 (28.9\%) \\
exim & 4.89 & 43 & 93 & 968 & 607 (62.7\%) & 88 & 121 (137.5\%) & 359 & 165 (46.0\%) \\
findutils & 4.6.0 & 28 & 52 & 821 & 554 (67.5\%) & 200 & 89 (44.5\%) & 326 & 65 (19.9\%) \\
grep & 2.25 & 19 & 28 & 460 & 264 (57.4\%) & 38 & 19 (50.0\%) & 113 & 52 (46.0\%) \\
httpd & 2.4.25 & 248 & 546 & 2,800 & 2,338 (83.5\%) & 1,332 & 483 (36.3\%) & 3,915 & 794 (20.3\%) \\
lighttpd & 1.4.45 & 27 & 108 & 899 & 524 (58.3\%) & 228 & 40 (17.5\%) & 830 & 221 (26.6\%) \\
ncurses & 6.0 & 46 & 77 & 1,835 & 1,045 (56.9\%) & 156 & 273 (175.0\%) & 969 & 397 (41.0\%) \\
nginx & 1.10.1 & 84 & 290 & 1,299 & 977 (75.2\%) & 610 & 319 (52.3\%) & 5,977 & 3,512 (58.8\%) \\
sed & 4.2.2 & 1 & 1 & 213 & 140 (65.7\%) & 2 & 0 (0.0\%) & 2 & 0 (0.0\%) \\
tar & 1.28 & 46 & 86 & 1,166 & 730 (62.6\%) & 141 & 166 (117.7\%) & 1,008 & 754 (74.8\%) \\
util-linux & 2.27.1 & 53 & 75 & 3,143 & 1,681 (53.5\%) & 211 & 177 (83.9\%) & 1,060 & 643 (60.7\%) \\
zlib & 1.2.8 & 5 & 14 & 152 & 108 (71.1\%) & 5 & 0 (0.0\%) & 13 & 0 (0.0\%) \\

  \bottomrule
  \end{tabular}
\end{table*}

In order to establish a baseline for the accuracy of type signatures, we have to
distinguish valid and invalid targets. Because of the imprecision of points-to
analysis, no automated analysis can perform sound and precise determination of
valid and invalid targets, so we first manually analyze the JSON output and
label targets as valid or invalid using careful inspection of the source code.
This establishes the baseline to which we can compare RAP's type signatures.
Note that the manual labeling is not because of the limitations in the analysis
of this effort; rather, it is a fundamental limitation of points-to analysis.
Because no automated analysis can establish sound and precise points-to
analysis, comparing RAP's targets with any other automated analysis would be
meaningless; there would be no basis to believe one as the ground truth.
Moreover, if such a precise automated analysis existed, points-to analysis-based
CFI could have been implemented precisely. However, we know that points-to
analysis is imprecise, and in fact, the imprecisions have been shown in the
previous work~\cite{ccs2015jujutsu}. That is why we manually label targets. The
rest of our analysis is automated.

Table~\ref{tab:over_approximation} illustrates the statistics for the analyzed
applications, among which the number of function pointer signatures defined by
the program and the number of all locations from which these function pointers
are being called (call sites). For example, observe that applications such as
\texttt{httpd}, heavily use function pointers due to their modular design.

Furthermore, we extract all functions defined in the programs, as well as the
functions for which RAP generates a type signature (hash). Interestingly, RAP
generates no hash for 34.6\% of the functions, on average, across all programs.
One reason is the fact that RAP does not generate a hash for \texttt{static}
functions which cannot be invoked by external callers and are not called
indirectly locally. Our analysis indicates that a large fraction of functions
are \textit{not} called indirectly. Specifically, only 15.2\% of functions are
also valid targets on average across all programs. In other words, about 50.2\%
of functions on average have a hash generated for them, while they are not
called indirectly at all. This, consequently, increases the hash collisions and
creates further opportunities for an attacker to call these functions using
irrelevant function pointers with the same signature. As mentioned before, it is
not uncommon for practical implementations to include non-address taken
functions into the target sets. We observed this behavior in both RAP and LLVM-
CFI implementations.

The important factor for estimating the impact of type collision is determining
the number of functions that are not valid targets of function pointers but can
be indirectly called through one of the function pointers in the program. The
eighth and tenth columns of Table~\ref{tab:over_approximation} show the number
of invalid target functions and invalid indirect calls possible in the program
because of type collisions.

These numbers underline the scope of the weakness created by type collisions. In
Nginx alone, for example, 3,512 function pointers can invalidly point to 319
functions which are never intended to be indirect targets just because they
happen to share the same type signature. Note that in applications that rarely
use function pointers, such as \texttt{base-passwd} and \texttt{sed}, the number
of possible corruptible indirect calls because of type collision is zero, but
these applications are the exceptions rather than the rule. In our analysis, any
application that contains more than tens of function pointers present abundant
opportunities for function pointer corruptions that respect the RTC.

\subsection {Gadget Distribution}

\begin{table}[t]
  \caption{Gadget distributions.}
  \label{tab:gadget_table}
  \setlength{\tabcolsep}{4pt}
\small
  \centering
  \begin{tabular}{llrrr}

  \toprule

  \textbf{App} &
  \textbf{Version} &
  \textbf{\gadgetone} &
  \textbf{\gadgettwo} &
  \textbf{\gadgetthree}
  \\

  \midrule
  nginx & 1.10.1 & 8 & 6 & 1 \\
  httpd & 2.4.25 & 40 & 19 & 5 \\
  lighttpd & 1.4.45 & 8 & 29 & 6 \\
  exim & 4.90 & 16 & 32 & 7 \\

  \bottomrule
  \end{tabular}
\end{table}

Now that we know type collisions can result in numerous opportunities for
control redirection, we shift our focus to counting the gadgets necessary for an
exploit. We use the algorithms described in Section~\ref{sec:attack}.
Table~\ref{tab:gadget_table} shows the number of gadgets in four popular web and
mail servers. Any combination of these gadgets could be a new invocation chain as described in section 3.3 and 3.4. However, more inter-procedural  analysis might be needed to determine controllable {\gadgettwo}s.
For these results, we limited our analysis only to the programs
themselves and not the linked libraries because this provides a more accurate
result, and avoids double counting gadgets in overlapping sets of linked
libraries. Moreover, as described earlier, the {\gadgetthree}s can be many
different targets depending on the exact goal of the payload (\eg modifying
target configurations, disabling W $\oplus$ X, running a shell script, \etc).
For this analysis, we chose a general, yet powerful type of {\gadgetthree}s,
namely those that allow arbitrary execution via spawning a malicious shell (\eg
\textit{exec} family or \textit{system}). We also chose {\gadgetone}s based on
the path to the {\gadgetthree}s from the list of invalid function targets.

As can be observed from the table, there are many gadgets available in these
applications. In fact, Nginx has the lowest number of gadgets among the four
servers analyzed, but as demonstrated, we could successfully hijack its control
and launch a malicious shell. In our exploit, we used four gadgets from the 15
available in Nginx.

Linked libraries provide numerous other opportunities for malicious control
hijacking in the face of RTC. For the sake of simplicity and accuracy, our gadget
counts in Table~\ref{tab:gadget_table} do not include the gadgets from linked
libraries, but for the sake of completeness, we now evaluate the impact of
linked libraries, primarily the ubiquitous library in Linux applications and
servers, Libc.

\subsection{Libc}

For complete protection, the linked libraries must also be protected with RTC.
However, counter-intuitively, here we show that protecting linked libraries with
RTC significantly increases the number of opportunities for the attacker.

To evaluate the impact of linked libraries, we compile the applications listed
in Table~\ref{tab:over_approximation} this time with
\texttt{glibc}\footnote{https://www.gnu.org/s/libc/}, and recount the number of
invalid indirect targets and invalid edges (indirect calls), introduced by the
additional collisions. Figure~\ref{fig:libc:targets} shows the number of targets
in the analyzed applications with and without \texttt{glibc}. We observe that
linking \texttt{glibc} significantly increases the number of targets that have
collisions with the function pointers in the applications, thus magnifying the
opportunities for an attack. The new collisions open new paths for attackers to
transfer control from the function pointers in the application to the functions
inside \texttt{glibc} while respecting RTC. For example, in \texttt{coreutil},
there are 297 new target functions to which the execution can be transferred,
and in \texttt{grep}, 291 target functions are added to the existing gadgets.

More important than potential targets, however, are the additional edges.
Figure~\ref{fig:libc:edges} illustrates the additional edges (indirect calls)
allowed in the applications as a result of collisions introduced by
\texttt{glibc}. We observe that the number of the new edges is much higher than
the number of new target functions in \texttt{glibc}. This is because different
function pointers can call all of the colliding target functions, thus growing
the number of possible edges multiplicatively.

\subsection{Type Checking vs. Points-to Analysis}
 
Previous attacks such as Control Jujutsu~\cite{ccs2015jujutsu} describe the
importance of precise CFG. Even though our focus is evaluating RTC-based CFI
variants, we posit that it can be useful to compare the over-approximation in
RTC with points-to analysis-based CFI. Recall that the over-approximation in
points-to analysis-based CFI is because of imprecise points-to analysis, while
in RTC-based CFI, it is because of type collisions.

In order to compare the two, we calculate the number of invalid edges allowed
because of their imprecisions. As a baseline for points-to analysis-based CFI,
we use SVF~\cite{ye2014region,sui2016svf}, which is the state-of-the-art in flow
sensitive points-to analysis. We again used manual labeling and source code
inspection to identify valid and invalid edges. Table~\ref{tab:rtc_vs_pta} shows
the number of total and invalid edges in both type checking and points-to
analysis. The base column shows the ground truth (manual labels). The results
provide no clear advantage for one or the other approach in terms of the number
of targets. In 4 of the 14 programs, RTC is less accurate than points-to
analysis, while in 8 of them, it is the opposite. However, we observe that flow
sensitive points-to analysis is not always possible. For example, for Nginx and
Httpd, SVF was not able to finish the analysis process, and in fact crashed
after 5 hours (showed by a dash in the table). On the other hand, RTC can be
applied in large code bases more easily. This experiment suggests that RTC is 
a more practical solution which offers almost the same security guarantees for
large real-world programs.

\begin{table}[t]
  \caption{Invalid indirect calls added to programs because of type collisions and imprecise points-to analysis.}
  \label{tab:rtc_vs_pta}
  \setlength{\tabcolsep}{4pt}
\small
  \centering
  \begin{tabular}{lrrrrrr}

  \toprule

  \makecell{\multirow{2}{*}{\textbf{App}}} &
  \makecell{\multirow{2}{*}{\textbf{Base}}} &
  \multicolumn{2}{c}{\textbf{Type Checking}} &
  \multicolumn{2}{c}{\textbf{Points-to Analysis}} \\

  \cmidrule[0.5pt](lr){3-4}
  \cmidrule[0.5pt](lr){5-6}

  &
  &
  \textbf{Total} & \textbf{Invalid} &
  \textbf{Total} & \textbf{Invalid} \\

  \midrule

  base-passwd & 0 & 0 & 0 (0.0\%) & 0 & 0 (0.0\%) \\
  coreutils & 213 & 291 & 78 (26.8\%) & 308 & 198 (64.3\%) \\
  e2fsprogs & 557 & 861 & 304 (35.3\%) & 42 & 15 (35.7\%) \\
  exim & 107 & 212 & 105 (49.5\%) & 169 & 99 (58.6\%) \\
  findutils & 237 & 279 & 42 (15.1\%) & 448 & 231 (51.6\%) \\
  grep & 54 & 105 & 51 (48.6\%) & 108 & 60 (55.6\%) \\
  httpd & 2,126 & 2,870 & 744 (25.9\%) & - & - \\
  lighttpd & 327 & 442 & 115 (26.0\%) & 1,096 & 938 (85.6\%) \\
  ncurses & 291 & 558 & 267 (47.8\%) & 507 & 238 (46.9\%) \\
  nginx & 1,276 & 2,287 & 1,011 (44.2\%) & - & - \\
  sed & 2 & 2 & 0 (0.0\%) & 2 & 0 (0.0\%) \\
  tar & 208 & 664 & 456 (68.7\%) & 360 & 167 (46.4\%) \\
  util-linux & 311 & 943 & 632 (67.0\%) & 596 & 465 (78.0\%) \\
  zlib & 10 & 10 & 0 (0.0\%) & 10 & 4 (40.0\%) \\

  \bottomrule
  \end{tabular}
\end{table}

\section{Discussion}
\label{sec:disc}

There are further practical challenges of implementing RTC for real-world
applications. For the sake of completeness, we review them briefly in this
section.

\subsection {Type Diversification}

As it was shown in our study, the main source of problem in RTC is type
collision. RAP proposes a type diversification technique in order to generate
unique types. Diversification of colliding types, in such a way that each
function pointer only shares type signature with its true target would reduce
the type collision problem. By this technique, the attacker will not be able to
find {\gadgetone} and therefore the attacks like \na~ could be mitigated from
the beginning. However, such a diversification, unlike generic RTC, requires a
precise points-to analysis to establish the exact target of each function
pointer. As discussed earlier, precise points-to analysis is shown to be hard,
and the imprecisions are shown to be exploitable~\cite{ccs2015jujutsu}. In other
words, type diversification re-introduces the challenge faced by points-to
analysis-based CFI, namely the imprecision of points-to analysis, to RTC.

\subsection{Separate Compilation}

Large code bases are often compiled in separate compilation units at different
times. This allows easier collaboration and debugging in large projects.
However, separate compilation further complicates type diversification. Since
not all source files are available during each compilation step, the colliding
types to diversify are not known to RTC. If each compilation unit is diversified
independent of the other units, false positives will be introduced and execution
will halt whenever function pointer from a compilation unit benignly calls a
function in a different compilation unit (because their diversified types are
very likely different). Consequently, proper type diversification in the
presence of separate compilation units requires additional meta data and tracking
mechanisms to avoid breaking functionality.

\subsection{Mismatch types}

\begin{figure}
  \begin{lstlisting}
  typedef ngx_int_t (*ngx_output_chain_filter_pt)
    (void *ctx, ngx_chain_t *in);
  
  static ngx_int_t ngx_http_charset_body_filter
    (ngx_http_request_t *r, ngx_chain_t *in)
  \end{lstlisting}
\vspace{-.2in}
    \caption{Type mismatch between the \texttt{ngx\_output\_chain\_filter\_pt} function
    pointer and its target \texttt{ngx\_http\_charset\_body\_filter}.}
    \label{fig:type_mismatch}

\end{figure}

The main assumption in RTC is that the types of caller and callee are the same.
However, this assumption is not always true. For example in real-world
applications there are cases that some part of the function pointers are
\textit{void*} and these function pointers are able to point to any other
pointer types such as \textit{int*} and \textit{long*}. According to RTC, this
is a mismatch. However, in fact, this feature makes the programs more flexible.
Figure~\ref{fig:type_mismatch} shows an example of this type mismatch between
the function pointer with its valid function target in Nginx.

In order to address this challenge, there are two possible approaches. The first
approach is to cast all the \textit{void*} to the proper types in order to
prevent any mismatch. This approach, used by RAP, requires large modifications
in the source code, which makes it time-consuming. The second approach is to
generalize \textit{void*} to other types and allow matching of \textit{void*} to
any pointer type. Although this method does not require source code
modifications, it makes the RTC more relaxed and creates more opportunity for
attacks.

\subsection{Support for Assembly Code}

In many low-level libraries, there are portions of the code that are written in
in-line assembly. Prominently, \texttt{glibc} implements system calls in assembly.
Automated annotation of assembly with type information is hard. A permissive
policy for assembly code could create additional exploitation opportunities,
while a restrictive one crashes benign applications. Further work is necessary
in this area to safely handle assembly code.

\section{Related work}

The related work in the area of memory attacks and defenses is vast. For broader
treatment of the related work, we refer the reader to surveys and
systematization of knowledge papers in this
area~\cite{EternalWar,diversity,cfi_survey}. Here we limit our discussions to
closely related efforts. Control Jujutsu attack~\cite{ccs2015jujutsu} bypasses
points-to analysis-based CFI using the additional edges introduced to a CFG
because of the imprecision of points-to analysis. Control Jujutsu further
demonstrates that certain coding practices such as recursions and modular design
exacerbate the imprecision of context-sensitive and flow-sensitive points-to
analysis. Control-Flow Bending~\cite{usenixsec2015cfb} also attacks CFI. It
assumes a fully-precise CFG, but uses versatile functions such as
\texttt{printf()} as its mechanism. Our attack is not reliant on any specific
function like that. Furthermore, a CFB attack is prevented by the modern
\texttt{Fortify Source} option in libc that stops ``\%n'' format string
attacks~\cite{percentn}. Counterfeit Object Oriented-Programming
(COOP)~\cite{COOP} is another attack on CFI. COOP exclusively relies on C++
virtual functions for its exploitation technique.

\section{Conclusion}

In this work, we evaluated the effectiveness of CFI techniques based on Runtime
Type Checking (RTC). We examined RTC from security and practicality
perspectives. We showed that while direct type collisions between corruptible
forward edges and target functions are rare, type collisions with other
functions can be exploited in a nested fashion to implement an attack (\na). We
further evaluated the prevalence of opportunities for such attacks and showed
that both the type collisions and the gadgets necessary for {\na} attacks are
abundantly found in many real-world applications. We also compared the
imprecisions of RTC and points-to analysis techniques and found that their
strength is heavily dependent on the code base. Our findings indicate that,
while RTC is a practical defense that can complicate exploitation, on its own,
it is not sufficient to prevent control-hijacking memory corruption attacks.

\section*{Acknowledgments}

We thank our shepherd, Sangho Lee, and the anonymous reviewers for their helpful comments. This material is based upon work supported by the National Science Foundation under Grant No. 1409738. This research was also sponsored by the U.S. Department of the Navy, Office of Naval Research, under Grant No. N00014-15-1-2948.

\bibliographystyle{ACM-Reference-Format}
\bibliography{paper}

\end{document}